\documentclass[a4paper]{jpconf}
\usepackage{graphicx}
\usepackage{ulem}
\usepackage{color}
\newcommand{\dg}{^{\dagger}}
\newcommand{\bk}{{\bf k}}
\newcommand{\bR}{{\bf R}}

\newlength{\fight}
\newcommand{\fg}[3]
{
\begin{figure}[ht]
\vspace*{-0cm}
\[
\includegraphics[width=\fight]{#1}
\]
\vskip -0.2cm
\caption{\label{#2}
\small#3
}
\end{figure}}
\begin{document}
\title{Origin of the Large Anisotropy in the $\chi_3$ Anomaly in $URu_2Si_2$}

\author{P. Chandra$^1$, P. Coleman$^{1,2}$ and R. Flint$^3$}

%\address{Production Editor, \jpcs, \iopp, Dirac House, Temple Back, Bristol BS1~6BE, UK}

\address{$^1$ Center for Materials Theory, Department of 
Physics and Astronomy, Rutgers University, 
Piscataway, NJ 08854}

\address{$^{2}$Department of Physics, Royal Holloway, University
of London, Egham, Surrey TW20 0EX, UK.}

\address{$^3$ 
Department of Physics, Massachusetts Institute for Technology,
77 Massachusetts Avenue, Cambridge, MA 02139-4307}

\ead{pchandra@physics.rutgers.edu}

\begin{abstract}
Motivated by recent quantum oscillations experiments on $URu_2Si_2$,
we discuss the microscopic origin of the large anisotropy observed
many years ago in the anomaly of the nonlinear susceptibility in this
same material.  We show that the magnitude of this anomaly emerges
naturally from hastatic order, a proposal for hidden order that is a
two-component spinor arising from the hybridization of a non-Kramers $\Gamma_5$
doublet with Kramers conduction electrons.  A prediction is made for the
angular anisotropy of the nonlinear susceptibility anomaly as a test
of this proposed order parameter for $URu_2Si_2$.
\end{abstract}

The character of the long-range ``hidden order'' (HO) in the heavy fermion material $URu_2Si_2$ continues
to fascinate and elude theorists and experimentalists \cite{Mydosh11}.  Despite a large entropy released
at the HO transition temperature ($T_0 = 17.5 K$) accompanied by many sharp
thermodynamic anomalies there\cite{Mydosh11,Palstra95,Miyako91,Ramirez92},  
there is no consensus on  the important local degrees of freedom 
and more importantly how they form the hidden order state.  

A revisit to the nonlinear susceptibility ($\chi_3$) and its sharp anomaly at $T_0$ \cite{Miyako91,Ramirez92}
may provide insight into the mysterious nature of the hidden order.  Originally the behavior of this
mean-field-like anomaly inspired a density-wave description. A simple Landau theory, $f = f(T - T_c(B^2))$, 
was constructed that yielded the relation
\begin{equation}
\Delta \left(\frac{c_v}{T}\right) \Delta \chi_3 = 12 \left[ \Delta \left(\frac{\partial \chi}{\partial T}\right)\right]^2
\end{equation}
that was tested successfully experimentally\cite{Ramirez94}, suggesting underlying itinerant order; here we recall
that $M = \chi_1 B + \frac{\chi_3}{3!} B^3$ where $M$ is the magnetization and $\chi_1$ and $\chi_3$ are the linear
and nonlinear susceptibilities respectively.  However the
large c-axis anisotropy of this $\chi_3$ anomaly was then emphasized and ascribed to local Ising f-moment 
behavior \cite{Santini94,Santini98}.  The appropriate Landau theory is then 
\begin{equation}
f(T,B_z) = [\alpha(T_c - T) - \eta_z B_z^2] |\Psi|^2 + \beta |\Psi|^4 
\end{equation}        
where $\Psi$ is the hidden order parameter; here a key question arises:  what is the microscopic origin of the large anisotropy coefficient $\eta_z$?

The anisotropic field-response of the U-ions in $URu_2Si_2$ is most naturally understood 
if their low-energy configurations are $5f^2$, and this point has been noted by several 
authors\cite{Santini94,Santini98,Amitsuka94,Ohkawa99,Flint12}.  Such strong Ising anisotropy was also
observed in dilute $U_xTh_{1-x}Ru_2Si_2$, thus indicating that it is a key property of
the U ions \cite{Amitsuka94}.  It was originally
described by quadrupolar ordering of the local f-moments\cite{Santini94,Santini98,Ohkawa99}, but 
unfortunately recent resonant X-ray measurements see no 
signature for such multipolar order\cite{Amitsuka10,Walker11}.
Furthermore there is experimental evidence from 
scanning tunnelling microscopy\cite{Schmidt10,Aynajian10},
and de Haas-van Alphen (dHvA) and 
Shubnikov-de Haas (SdH)\cite{Ohkuni99,Jo07,Hassinger10,Altarawneh11} 
measurements for itinerant behavior below $T_o$, so a 
local-moment scenario cannot capture the full picture at the transition. 

Recent quantum oscillations provide a clue towards resolving this
dilemma, as they reveal that the quasiparticles deep in the HO phase
exhibit a large Ising anisotropy\cite{Altarawneh11}.  More
specifically their Zeeman splitting depends only on the c-axis
component of the magnetic field, $\Delta E = g(\theta) \mu_B B$ with a
g-factor $g(\theta) = g \cos \theta$.  The size of the measured g-factor
anisotropy is resolution limited and
exceeds thirty, corresponding to an anisotropy of the
Pauli susceptibility in excess of 900 that is too large to be attributed simply
to spin-orbit coupling.  These experiments suggest that the anisotropy
of the U $5f^2$ moments is transferred to the mobile quasiparticles in
the HO phase via hybridization.  Angle-dependence of the Pauli-limited
critical field of the superconducting
state\cite{Brison95,Altarawneh12} confirms the g-factor anisotropy;
furthermore the presence of Ising quasiparticles in a superconductor
with $T_c \sim 1.5 K$ indicate that the U $5f^2$ configuration is
degenerate to within an energy resolution of $g\mu_BB \sim 5K$.  The
tetragonal symmetry of $URu_2Si_2$ protects the non-Kramers doublet
$\Gamma_5$ that is quadrupolar in the basal plane and magnetic along
the c-axis, and it has been proposed as the origin of the magnetic
anisotropy in both dilute and dense
$URu_2Si_2$\cite{Amitsuka94,Ohkawa99,Altarawneh11,Chandra12}.

Conventionally in heavy fermion materials, hybridization involves
valence fluctuations between a ground-state Kramers doublet and an
excited singlet; it develops via a crossover, leading to mobile heavy
quasiparticles.  However if the ground-state is a non-Kramers doublet,
the Kondo effect will involve an excited Kramers
doublet\cite{Amitsuka94}.  Hybridization now carries spin and
coherence; its development breaks time-reversal and spin rotational
invariance at a true phase transition.  Furthermore because it mixes
states of half-integer (non-Kramers doublets) and integer (conduction
electrons) it breaks Kramers parity and thus also breaks double-time
reversal\cite{Chandra12}.  In $URu_2Si_2$ optical\cite{Bonn88} and
tunnelling probes\cite{Schmidt10,Aynajian10} indicate that
hybridization develops abruptly at the HO transition, and thus we have
proposed that the hybridization is a two-component order parameter
(``hastatic'' order) that transforms as a spinor\cite{Chandra12}.
\fight=\textwidth
\fg{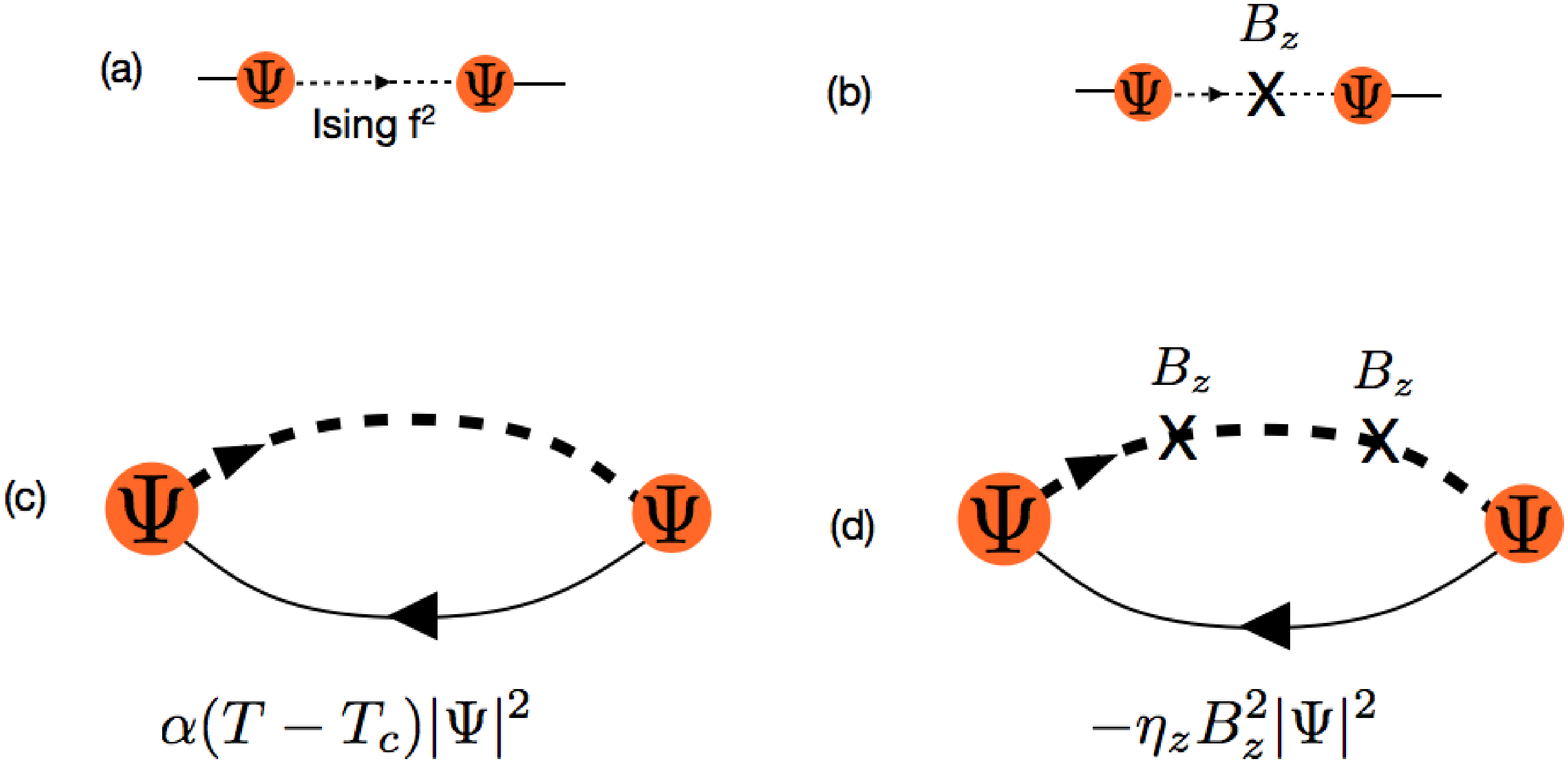}{fig1}{Showing (a) resonant Kondo scattering via 
hybridization off non-Kramers $5f^{2}$ doublet, (b) coupling to
magnetic field via f-state is purely Ising like (c) zero field
quadratic contribution to the Landau energy (d) Ising field-dependent 
contribution $-B_{z}^{2}\Psi^{2}$ term in the Landau theory.
}

Let us briefly digress to discuss the microscopic valence fluctuations
of a non-Kramers doublet. 
The valence fluctuation physics of a non-Kramers doublet are
described by an Bolech-Andrei model\cite{Bolech02} (ABM) $H=\sum_{\bk
}\epsilon_{\bk }c\dg_{\bk \sigma }c_{\bk \sigma }+ \sum_{j}H_{ABM}(j)$,
where the valence fluctuations and atomic physics of each $U$ ion are
described by 
\begin{equation}\label{}
H_{ABM}(j) = V \biggr[
c\dg _{\sigma \alpha } (j)X_{\sigma \alpha } (j)
 + {\rm H.c}\biggr] +
\epsilon X_{\sigma \sigma } (j)
\end{equation}
where the $X_{\sigma \alpha }= \vert 5f^{3},\sigma \rangle \langle 5f^{2},\alpha \vert 
$ is the Hubbard operator between  the Kramers and non-Kramers
state, $X_{\sigma \sigma } = \vert 5f^{3},\sigma \rangle \langle
5f^{3},\sigma \vert $ is the projection onto the excited Kramer's
doublet, and
$\epsilon= E[5f^{3}]-E[5f^{2}]$ is the energy difference
between the $5f^{3}$ excited Kramers doublet and the $5f^{2}$
non-Kramers ground-state. The operator $c\dg _{\sigma \alpha }
(j)$ creates a conduction {\sl hole} with channel and spin index
$\alpha $ and $\sigma $ at site $j$.   
Valence fluctuations out of a  non-Kramer's doublet involve at least
two different crystal symmetry channels, the $\Gamma_{7}$ and
$\Gamma_{6}$ channels. The ABM 
results from the projection of an Anderson model into the low-lying
$5f^{2}$, $5f^{3}$ subspace. $H_{ABM} (j)= {\cal P}H_{Anderson}
(j){\cal P}$.  Typically the $c\dg_{\sigma \alpha }$
is expanded in terms of the two valence fluctuation channels 
$\lambda = 1,2 \equiv \Gamma_{6,7}$ as follows
\begin{equation}\label{}
c\dg_{\sigma \alpha }= \sum_{\lambda=1,2}c\dg_{\lambda \sigma '}\Gamma^{\sigma '}_{\sigma \alpha } (\lambda),\qquad 
\Gamma^{\sigma'}_{\sigma \alpha } (\lambda) = \langle 5f^2\alpha\vert
f_{\lambda\sigma '}\vert 5f^{3}\sigma \rangle .
\end{equation}
where $c\dg_{\lambda \sigma }= \sum_{\bk, \sigma' }\gamma^{\lambda }_{\sigma
\sigma '} (\bk ) c^{\dg }_{\bk \sigma }e^{-i \bk \cdot {\bR}_{j}}
$
defines the spin-orbit coupled Wannier states of the conduction holes
in terms of the crystal-field form-factors $\gamma_{\sigma
\sigma '}^{\lambda} (\bk )$ and plane wave conduction states.

If we represent $X_{\sigma \alpha }= \hat \Psi \dg_{\sigma }f_{\alpha
}$ using a slave boson decomposition, 
where $\Psi\dg _{\sigma }$ is a slave boson describing the excited
Kramers' doublet $\hat \Psi \dg_{\sigma }\vert 0\rangle \equiv  \vert 5f^{3},
\sigma \rangle $ while $f \dg_{\alpha }$ is an Abrikosov
pseudo-fermion 
describing the low-lying non-Kramer's doublet $f \dg_{\alpha }\vert
0\rangle \equiv \vert 5f^{2},\alpha \rangle $. 
Hastatic order corresponds
to the condensation of the slave boson, 
\begin{equation}\label{}
X_{\sigma \alpha }\rightarrow \langle \hat \Psi \dg_{\sigma }\rangle
f_{\alpha } \equiv \Psi^{*}_{\sigma }f_{\alpha }
\end{equation}
The condensed slave boson defines the two-component Hastatic order
parameter, 
giving rise to hybridization terms of the form shown in
Fig. {\ref{fig1}} (a). Although this term looks like a conventional
hybridization in a heavy fermion system, it breaks time reversal
symmetry in a new way, by mixing an
integer spin non-Kramers doublet with a half-integer spin conduction
electron. 
The coherent, broken symmetry physics of the HO phase is 
is determined by the hybridization Hamiltonian
\begin{equation}\label{}
H^{*}(j) = V \biggr[
\Psi^{*}_{\sigma }c\dg _{\sigma \alpha } (j) 
f_{\alpha }(j)
+ {\rm H.c}\biggr] +
\epsilon |\Psi |^{2}
\end{equation}

Let us now return to the anomaly in $\chi_3$.  In the field-dependent
Landau theory of hastatic order, we get the first
temperature-dependent term (Fig. 1 a) from the hybridization of the
conduction electrons and the Ising f-moments. We can obtain the Landau
theory from a Feynman diagram expansion of the Free energy in powers of
$\langle \Psi \rangle $, as illustrated in Fig. 1. 
Applying a magnetic
field, the conduction electrons couple to it isotropically, whereas
the non-Kramers doublet couples linearly only to $B_z$(Fig. 1b).
Thus the
resulting field-dependent Landau theory, depicted schematically in Fig1c and Fig1d,
is
\begin{equation}
f[\psi] = 
[\alpha(T-T_{c}) - \eta_z B_z^2 -\eta_\perp B_\perp^2]\Psi^2 
+ \beta \Psi^4
\end{equation}
where $B_z = B \cos \theta$ and $B_\perp = B \sin \theta$. 
Minimizing the free energy with respect to $\Psi $, we obtain
\begin{equation}\label{}
f =  - \frac{[\alpha(T_{c}-T) + B^{2} (\eta_z \cos^2\theta + \eta_\perp\sin^2\theta)]^{2}}{4\beta }
\end{equation}
from which
\[
\Delta \chi_{3} = - \frac{\partial^{4}f}{\partial B^{4}} = 
\frac{6}{
\beta }
(\eta_z \cos^2\theta + \eta_\perp\sin^2\theta)^{2}
\]
We can calculate $\eta_z$ and $\eta_\perp$ within a
two-channel Anderson lattice model of hastatic order \cite{Chandra12},
and we find that $\frac{\eta_\perp}{\eta_z} = \frac{T_0^2}{D^2}$ where
$T_o$ and $D$ are the hidden order transition temperature and the
bandwidth respectively.  Taking a conservative $\frac{T_0}{D} \sim
\frac{1}{30}$, leads to an anisotropy of $10^3$ in
$\frac{d\chi_1}{dT}$ and thus a factor of $10^6$ in $\Delta\chi_3$. Of
course our model is simplified and there will be many {material}
features (e.g. f-electron contributions to $\eta_z$ {arising from fluctuations to excited states}) that will reduce
the anisotropy.  The crucial point here is that the anisotropy in the
$\chi_3$ anomaly will be several orders of magnitude greater than the
single-ion anisotropy in $\chi_1$,
and that $\chi_3 \propto \cos^4 \theta$
where $\theta$ is the angle from the c-axis.  This enormous Ising
anisotropy in the anomaly of $\chi_3$ is a natural consequence
of hastatic order and is one of its many testable consequences. 
Careful new measurements to determine the angular dependence of 
$\Delta \chi_3$ would verify the Ising nature of the quasiparticles.

In conclusion we have shown that hybridization between half-integer
conduction electrons and spin-half non-Kramers doublets provides a
natural microscopic explanation for the large anomaly observed in the
nonlinear susceptibility anomaly\cite{Miyako91,Ramirez92} in
$URu_2Si_2$.  Here the f-moments couple only to the c-axis component
of the applied field and, via hybridization, transfer their Ising
anisotropy to the conduction electrons.  These mobile Ising
quasiparticles form Cooper pairs at lower temperatures, and 
the observable consequences of their hastatic nature in the superconducting
state remain to be explored.

{\bf Acknowledgments.} 
We are grateful to helpful discussions with N. Andrei, N. Harrison, K. Haule,
G. Kotliar, Y Matsuda and J. Mydosh. We would
like to thank the Aspen Center for Physics, where this project was
started and then completed during two separate visits.
This work was supported under grants DMR 0907179 (P. Coleman),
NSF grant 1066293 (P. Coleman, P. Chandra and R. Flint) and a Simons
Foundation fellowship (R. Flint).


\noindent{\bf References}

\begin{thebibliography}{99}

\bibitem{Mydosh11} J.A. Mydosh and P.M. Oppeneer, {\sl Rev. Mod. Phys.} {\bf 83}, 1301 (2011).

\bibitem{Palstra95} T.T.M. Palstra et al, {\sl Phys. Rev. Lett.} {\bf 55} 2727 (1085).

\bibitem{Miyako91}  Y. Miyako et al, {\sl J. Appl. Phys.} {\bf 70} 5791 (1991).

\bibitem{Ramirez92} A.P. Ramirez et al. {\sl Phys. Rev. Let.} {\bf 68}, 2680 (1992).

\bibitem{Ramirez94} A.P. Ramirez et al. {\sl Physica B} {\bf 199 \& 200}, 426 (1994).

\bibitem{Santini94} P. Santini and G. Amoretti, {\sl Phys. Rev. Lett.} {\bf 73}, 1027 (1994).

\bibitem{Santini98} P. Santini, {\sl Phys. Rev. B} {\bf 57}, 5191 (1998).

\bibitem{Amitsuka94} H. Amitsuka and T. Sakakibara, {\sl J. Phys. Soc. Japan} {\bf 63}, 736 (1994).

\bibitem{Ohkawa99} F.J. Ohkawa and H. Simizu, {\sl J. Phys. Cond. Mat.} {\bf 11}, L519 (1999).

\bibitem{Flint12} R. Flint, P. Chandra and P. Coleman, arXiv:1207.2433 (2012).

\bibitem{Amitsuka10} H. Amitsuka et al, {\sl J. Phys.: Conf. Series} {\bf 200}, 012007 (2010).

\bibitem{Walker11} H. Walker et al. {\sl Phys. Rev. B} {\bf Phys. Rev. B} {\bf 83}, 193102 (2011).

\bibitem{Ohkuni99} H. Ohkuni et al, {\sl Phil. Mag} {\bf 79} 1045 (1999).

\bibitem{Jo07}  Y. Jo et al., {\sl Phys. Rev. Lett.} {\bf 98} 166404 (2007).

\bibitem{Hassinger10} E. Hassinger et al.,{\sl Phys. Rev. Lett.} {\bf 105} 216409 (2010).

\bibitem{Altarawneh11}  M. M. Altarawneh et al, {\sl Phys. Rev. Lett.} {\bf 106} 146403 (2010).

\bibitem{Schmidt10}  A.R. Schmidt et al, {\sl Nature} {\bf 465}, 570 (2010).

\bibitem{Aynajian10} P. Aynajian et al., {\sl Proc. Natl. Acad. USA} {\bf 107} 10383 (2010).

\bibitem{Brison95} J.P. Brison et al., {\sl Physica C} {\bf 250}, 128 (1995).

\bibitem{Altarawneh12}  M.M. Altarawneh et al., {\sl Phys. Rev. Lett.}{\bf 108} 066407 (2012).

\bibitem{Chandra12} P. Chandra, P. Coleman and R. Flint, arXiv:1207.4828 (2012).

\bibitem{Bonn88} D.A. Bonn et al., {\sl Phys. Rev. Lett.} {\bf 61}, 1305 (1988).

\bibitem{Bolech02} C.Bolech and N. Andrei, {\sl Phys. Rev. Lett} {\bf 88}, 237206 (2002).

\end{thebibliography}
\end{document}